# Designing an Energy Efficient Framework for Data Gathering in Wireless Sensor Network

**Subhabrata Mukherjee**  *subhabrata.mukerjee.ju@gmail.com*

*Advanced Digital and Embedded Systems Lab,*

*Department of Electronics and Telecommunications Engg.,*

*Jadavpur University, Kolkata 700032, India*

**Anand Seetharam**  *anandsthrm@yahoo.co.in*

*Advanced Digital and Embedded Systems Lab,*

*Department of Electronics and Telecommunications Engg.,*

*Jadavpur University, Kolkata 700032, India*

**Abhishek Bhattacharyya**  *abhishek.bhattacharyyya@gmail.com*

*Advanced Digital and Embedded Systems Lab,*

*Department of Electronics and Telecommunications Engg.,*

*Jadavpur University, Kolkata 700032, India*

**Mrinal.K.Naskar**  *mrinalnaskar@yahoo.co.in*

*Advanced Digital and Embedded Systems Lab,*

*Department of Electronics and Telecommunications Engg.,*

*Jadavpur University, Kolkata 700032, India*

**Amitava Mukherjee**  *amitava.mukherjee@in.ibm.com*

*IBM India Pvt. Ltd., Salt Lake, Kolkata 700091, India*

**Abstract**

Wireless sensor network (WSN) is a collection of nodes which can communicate with each other without any prior infrastructure along with the ability to collect data autonomously and effectively after being deployed in an ad-hoc fashion to monitor a given area. One major problem encountered in data gathering wireless systems is to obtain an optimal balance among the number of nodes deployed, energy efficiency and lifetime as energy of nodes cannot be replenished. In this paper we propose first a scheme to estimate the number of nodes to be deployed in a WSN for a predetermined lifetime so that total energy utilization and complete connectivity are ensured under all circumstances. This scheme also guarantees that during each data gathering cycle, every node dissipates the requisite amount of energy, which thus minimizes the number of nodes required to achieve the desired network lifetime. Second, this paper has proposed a framework to conduct data gathering in WSN. Extensive simulations have been carried out in ns2 to establish the effectiveness of this framework.

**Keywords**— Wireless sensor network, data gathering cycle, node estimation, network lifetime, energy efficiency.

## I. INTRODUCTION

Recent advancements in various fields of electronics have enabled the development of inexpensive low power sensors with significant computational capability [1]-[3]. Applications of sensor networks vary widely from climatic data gathering, seismic and acoustic underwater monitoring to surveillance and national security. The sensor networks are required to transmit gathered data to the base station (BS) or sink. It is often undesirable or infeasible to replace or recharge sensors. Network lifetime is thus an important parameter for sensor network design.

In case of WSNs, the definition of network lifetime is application specific [4]. It may be taken as the time from inception to the time when the network becomes nonfunctional. A network may become non-functional when a single node or a particular percentage of nodes drains out. However, it is universally accepted that equal energy dissipation for equalizing the residual energy of the nodes is one of the keys for prolonging the lifetime of the network [4]. Since sensors are constrained by limited battery power [1]-[2], [5] and communication requires significant amount of energy as compared to computations [1], nodes must collaborate in an energy-efficient manner for transmitting and receiving data so that lifetime enhancement is ensured. LEACH protocol [1] presented a solution to this energy utilization problem where nodes are randomly selected to collaborate to form a small number of clusters and the cluster heads take turn in transmitting to the base station during a data gathering cycle. PEGASIS protocol [2] was a further improvement upon LEACH protocol where a chain of nodes is formed which take rounds in transmitting data to the base station. Schemes in [6], [7] were improvements on the above strategies. However one of the drawbacks of all these data gathering schemes was observed that all these schemes showed how the network lifetime is increased. But the analysis had been performed with an arbitrary number of nodes. An estimate of the number of nodes required to be deployed in the area under surveillance was missing. Although authors' earlier work in [8],[11] provided an estimate of the number of nodes to be deployed many other aspects were not addressed. Moreover the coverage provided by the network had also not been analyzed in earlier papers. Variable radio ranges have been considered to maintain complete connectivity among nodes which is difficult to obtain in practice as the networks are homogeneous. In practice many of these schemes may not be viable as in real life situations before any WSN is deployed some parameters would definitely be provided to the designer to meet certain requirements. One of the primary things that would be supplied is the network lifetime or in other words the time for which the customer wants the network to survive. The end user would then want an estimate of the number of nodes to be deployed to achieve the desired lifetime. A further demand in most cases would be that the deployment be energy efficient. In this paper we have considered these aspects and designed a sensor network to meet certain specified requirements.

The rest of the paper is arranged as follows. In Section II we formally state the problem dealt with in this paper. The energy dissipation model considered has been described in Section III. Section IV specifies a strategy to achieve a definite coverage fraction. A detailed description of the data gathering cycle has been presented in Section V while a methodology for finding out the node density has been laid down is described in Section VI. Section VII describes the framework for data gathering and a flowchart illustrating the working of WSN has been laid down. Section VIII discusses in brief the fault tolerance management when a node failure occurs. Section IX contains the simulation results while Section X concludes the paper.

## II. THE PROBLEM STATEMENT

Given a continuously changing environmental phenomenon (e.g., temperature, humidity etc.) to be monitored in a specified region, the problem is to propose an efficient scheme for a data gathering WSN which meets the following requirements.

I. The network survives for a specified time ($T_{life}$).

II. A sufficient coverage fraction β must always be achieved during a data gathering cycle.

III. The deployment scheme is energy efficient.

IV. The network must collect a specific number of data packets (say K) from the region during each data gathering cycle.

V. An estimate of the number of nodes required to be deployed is to be provided.

## III. THE ENERGY MODEL

Firstly, we formulate an energy dissipation model appropriate for a data gathering WSN. To take into account all activities performed by a node during each data gathering cycle of duration $T_d$ we consider five states for all the nodes – the sensing, transmitting, receiving, idle and sleep states. An energy dissipation model for radio communication similar to [2], [10] has been assumed. As a result the energy required per second for successful transmission ($E_{ts}$) is thus given by,

$$E_{ts} = e_t + e_d d^n \qquad (1)$$

where $e_t$ is the energy dissipated in the transmitter electronic circuitry per second to transmit data packets and $e_d d^n$ is the amount of energy required per second to transmit over a distance d and n is the path loss exponent (usually $2.0 \leq n \leq 4.0$). The distance d, must be less than or equal to the radio range $R_{radio}$, which is maximum inter-nodal distance for successful communication between two nodes. If $T_{1b}$ is the time required to successfully transmit a bit over a distance d then total energy to transmit a bit is

$$E_t = (e_t + e_d d^n) T_{1b} \qquad (2)$$

If $e_r$ is the energy required per second for successful reception and if $T_{2b}$ is the total time required by a sensor to receive a bit then the total energy to receive a bit is

$$E_r = e_r T_{2b} \qquad (3)$$

Similarly if $e_{id}$ is the energy spent per second by the nodes in the idle state and $T_3$ is the time spent in the idle state, then,

$E_{id}=e_{id}\ T_3$         (4)

In the same manner if $e_{sen}$ is the energy dissipated by the nodes per second for sensing the environment and if $T_4$ be the time spent in this state then we have,

$E_{sense}=e_{sen}T_4$         (5)

The remaining part of the data gathering cycle is spent in sleep state of duration $T_5$, where we assume that no energy is dissipated. Moreover under practical situation one can take $T_{1b}=T_{2b}$. We consider that an active node forms one data packet of length 'B' bits during a data gathering cycle. In our scheme we have also defined a performance-measuring parameter, namely the energy utilization ratio (η) that is expressed as the ratio of the total energy used by the network during its lifetime to the total energy supplied to the network at the time of deployment. Thus η = ($E_{used}/E_T$)*100% where $E_{used}$ is the total energy utilized by the network during its lifetime and $E_T$ is the total energy of the nodes at the time of deployment.

## IV. ACHIEVING THE REQUIRED COVERAGE FRACTION

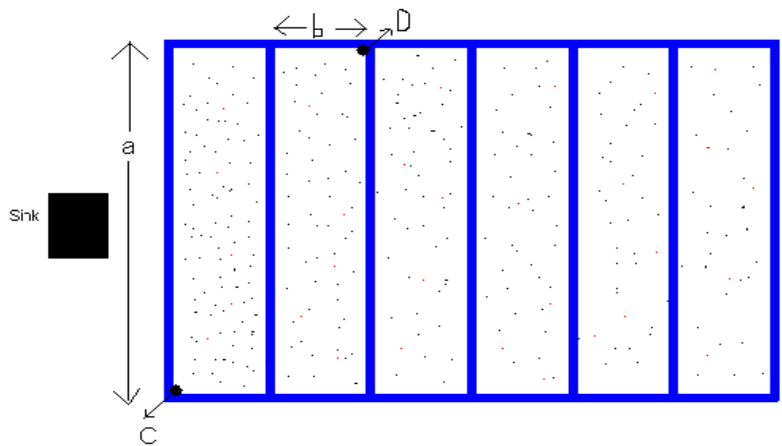

**Figure 1.** The total area to be monitored

We consider that the given area to be monitored is a rectangle of area A, as shown in Fig. 1. Let 'a' be its breadth and we divide its length into K equal divisions of width b so that one may collect the required number of data packets. So we have, A = abK. The section nearest to the sink is denoted as the 1st section while the last segment is the Kth section. In order to achieve a specific β during every data gathering cycle let us assume that the number of nodes in each segment required to be involved in data collection is 's'. We consider these nodes as the active nodes. As the standard motes can sense satisfactorily within a range of 10 m we take the sensing range (Rsense) as 10 m. This is the sensing range and is different from the radio range whose value is determined later. The value of s is obtained from (6). Let us consider a rectangle with length as 200 m and breadth equal to 60 m.
Let the given value of K be 20 and the required β be 90%.Therefore we have a=60 and b=10.We have from [11],

$fa=1-e^{-\lambda \pi R*R}$     (6)

(where fa is the coverage fraction, λ is the node density i.e number of nodes per square metre of the rectangle and R

is the sensing range).Putting fa=.9 for 90% coverage,R=10m we get λ=.008 nodes/sq metre.Thus within a strip of a=60 m and b=10 m,we have s=(.008 * 60 * 10)=4.8.Thus s=5 (rounding it to the nearest integer).

From the above discussion, we observe that the value of 's' can be obtained from equation 6. Although a specific case is cited here where s = 5, for the remainder of the paper we do all calculations considering the number of active nodes in each segment as 's' so that our discussions are valid in general.

## V. THE DATA GATHERING CYCLE

In this section we discuss the organization of the data gathering cycle in detail. For each active node the entire data gathering cycle is divided into the following stages. During the initial part of the data gathering cycle the active nodes are involved in sensing and in the collection of data. After this each node forms a single data packet of its own. Among 's' active nodes in every segment one of the nodes is selected as a leader. An open ended chain is formed starting from the leader in such a way that each node chooses its neighbor the nearest active node which is not already a part of the chain. The last node joining the chain transmits it packet to its neighbor which then fuses this packet with its own and transmits it to its other neighbor in the chain. Finally the leader transmits a single data packet of length 'B' bits. This is illustrated in the figure below. Let c0 - c4 be the 5 nodes selected as active nodes in a segment. Let c4 be chosen as the leader. A chain is formed as shown. c0 transmits its packet to c1 which fuses it with its own packet and in this way the data packet reaches the leader c4.

$$c0 \to c1 \to c2 \to c3 \to c4$$

Fig.2. The Chain Construction

The leader is rotated so that each active node is elected as a leader once in 's' data gathering cycles. This set of 's' cycles is called a data gathering round or simply 'Round'. The leaders now send these data packets to the sink by relaying them to their partners in the following section during the latter part of the data gathering cycle. The remaining part of the cycle is spent in the sleep state. Thus it is evident that as the nodes in the segments nearer to the sink have to deal with larger number of data packets as compared to those further away from the sink the energy expended by these nodes is greater. Let $E_o$ be the initial energy of each node deployed in the region. In the initial part of the data gathering cycle all the active nodes sense and collect data for a time $T_{sense}$ and thus we have $E_{sense} = e_{sen} T_{sense}$ for all the nodes. If $E_{i1}$ is the energy used up by an active node in the $i^{th}$ segment in a data gathering cycle when it is elected as a leader, then using (1-5) we have,

$$E_{i1} = BE_t(K-i+1) + BE_r(K-i+1) + E_{id1} + E_{sense}$$
$$= (e_t + e_d d^n)(K-i+1)T_1 + e_r(K-i+1)T_2 + e_{id}T_{idleil} + e_{sen}T_{sense} \quad (7)$$

where $T_{idleil}$ is the requisite amount of time for which leader in the $i^{th}$ segment must remain in the idle state during the data gathering cycle for successfully receiving and transmitting the requisite number of data packets.For simplification of analysis we neglect the energy required to fuse the data packets. Let $E_{i2}$ be the total energy spent by the active node in the remaining (s-1) data gathering cycle when it is not elected as leader. The detail analyisis of the value of $E_{i2}$ is shown later. Therefore the amount of energy spent by each active node in the $i^{th}$ section during a Round is

$$E_i = E_{i1} + E_{i2} \qquad (8)$$

where $E_i$ to be just the requisite amount of energy to be spent in one data gathering round.

## VI. DETERMINATION OF NODE DENSITY FOR ACHIEVING A DESIRED NETWORK LIFETIME

From discussions in Section V it is clear that all although active nodes in the same segment dissipate equal amount of energy during a Round the active nodes in the various segments would dissipate different amount of energy during the same time. Further it is a requirement that the network should last for a specific lifetime ($T_{life}$). From earlier discussions and (8) it can be easily concluded that the segments nearer to sink would require a larger number of nodes as compared to those further away from it. We now propose a deployment scheme so that $\eta = 100\%$ can be ensured. This would result in complete energy utilization and achieve the desired network lifetime with lesser number of nodes. Let $n_{di}$ be the number of nodes deployed per unit area in the $i^{th}$ segment to ensure $\eta = 100\%$. Thus for the desired network lifetime $T_{life}$ we have,

$$T_{life} = n_{di} a b T_d (E_o / E_i) \quad \text{for} \quad 1 \leq i \leq K \qquad (9)$$

where $T_d$ is the duration of the data gathering cycle. Thus from (9) one can calculate the node densities of the various segments. Now among all these sections we choose one of them (say the $K^{th}$ section) as a reference. This yields us the following relation.

$$n_{di} = n_{dK} \frac{E_i}{E_K} \quad \text{for} \quad 1 \leq i \leq K \qquad (10)$$

From (10) we observe that the node densities are directly proportional to the energy spent by an active node during a Round. As a result the total number of nodes in the $i^{th}$ segment is $N_i = n_{di} ab$. We round of $N_i$ to the next higher multiple of s. Finally the total number of nodes (N) that would be required to be deployed in the given area A can be approximately calculated by the relation stated in equation (11). This is because the following relation has been computed without taking into consideration the fact that the number of nodes in each section is rounded of to the next higher multiple of s.

$$N = ab\Sigma N_i \qquad (11)$$

Now from Section V, VI it is evident that if the values of $T_{idlei1}$ and $T_{idlei2}$, that can be determined for the nodes in different segments, are known then the value of the total number of nodes required to be deployed in the area A can be calculated. These values are calculated in Section VII.

## VII. AN EFFICIENT FRAMEWORK FOR DATA GATHERING

We have computed the number of nodes to be deployed in the various segments in Section VI. As the number of nodes in each section is much more than 's' at first begin with the nomenclature of the nodes. $N_i$ is the number of nodes to be deployed in the $i^{th}$ section. Each sensor node is named as S[i][j] where i denotes the section in which the node is deployed and j is the serial number of the node in that segment, $\forall$ i = 1 to K and $\forall$ j = 0 to $N_i$ -1. The nodes in each section are programmed to have their specific serial numbers and are then randomly scattered in their respective sections. We know that to achieve the required coverage fraction β during every data gathering cycle 's' nodes are required to be active in each segment. The rest of the nodes in each segment remain in the sleep state during this time. Now the selection of the s active nodes in every segment is done in the following way. In the first Round, nodes having serial numbers S[i][0] to S[i][s-1] are selected as the active nodes. As mentioned earlier each of these s active nodes becomes a leader once in a Round. After one Round, nodes having serial numbers between s and 2s-1 are entrusted with this duty. This process continues until all the nodes in a segment have been selected at least once. This complete process is called a 'Set'. After one entire 'Set' like this the first set of nodes becomes active again and participates in data gathering. Among the set of $N_i$ nodes in the $i^{th}$ section the ones that are going to participate in a data gathering cycle can be easily determined by the relation in (12). Let 'R' be the total number of rounds completed till then and let T be the number of sets completed. Among the active nodes in the $i^{th}$ section during a Round the one having the least serial number can be easily found out. The following relation (12) gives the serial number ($S_{no}$) of this particular node.

$(S_{no}) = sR - T N_i$ \hspace{2cm} (12)

So the serial numbers of the active nodes in the $i^{th}$ section vary from $S_{no}$ to $S_{no}$ + s -1. We know that during each cycle, in all segments one of the nodes is selected as the leader. The leader fuses the collected the data packets into a single data packet of length 'B' bits and transmits it to the leader in the next section. In every segment each of the active nodes becomes a leader once during a Round starting from the one having the lowest serial number among the selected nodes. The leader in the $(i+1)^{th}$ section transmits the data packet to the leader in the $i^{th}$ section and in this way all the K data packets reach the sink.

Certain features of IEEE 802.11 are not particularly suited for power sensitive wireless sensor networks. The power sensitive mechanism in 802.11 causes the unnecessary energy consumption due to the problems of an overhearing, a back-off time delay and possible packet collisions. Hence several MAC protocols have surfaced conforming to sensor network requirements, the S-MAC [9] being one of them. It is specially designed for event driven networks addressing burning issues of WSNs. It is different from IEEE 802.11 such that energy conservation and self configuration are primary goals rather than per node fairness and latency.
In this paper we outline an effective framework conforming to the data gathering system that we have proposed.
Equation 12 indicates which nodes are going to be active during a data gathering cycle. It is also possible to determine the leaders during each cycle. At the beginning of each data gathering cycle the sink synchronizes the activities of the various active nodes by sending a SYNC message to the nodes. As soon as the active nodes in different segments receive the SYNC message they simultaneously start the data gathering cycle. Initially all nodes devote themselves to data collection and after a time $T_{sense}$ the active nodes form their respective data packets. The nodes now transfer the data packets to their respective leaders via their neighbors as described in section V. In the proposed scheme we have neglected the transmission time required for the data packets to arrive at the receiver. However, the scheme can easily be modified if the transmission time delays are known. We now illustrate the working of the protocol by taking packet size 'B' bits and the data rate as 'D' bps. So the time for transmission and reception of a data packet is B/D secs. In the beginning active nodes in the odd segments form chains and transmit their data packet to the leader.

During this time the radio range ($R_{radio}$) is chosen as 'b'. This ensures that under no circumstances collisions take place. This is because even in the worst case scenario transmissions in segment 3 (say) cannot cause garbling up of information received at the receiver of a node stationed in segment 1 (say). During this time the nodes in the even segments remain in the sleep state. After a node transmits the requisite data packet to its neighbor it then goes into the sleep state. Thereafter these nodes in the odd segments go into the sleep state while those in the even segments transmit their data packets to their leaders. Now when each node transmits a data packet to its neighbor it would require B/D secs to transmit their data packet. So that withstanding clock drifts having magnitudes of nearly 0.2 ms per second [9] between two nodes do not result in any kind of error and packet loss we consider a total time duration of 2B/Dsecs which is twice the time required to successfully transmit or receive data packet. We know that an active node in a Round comprising of 's' data gathering cycles is selected as a leader once. So for the remaining 's-1' data gathering cycles in which it is not elected as leader the amount of energy it spends is

$$E_{i2} = (B/D)*(s-2)(s-3)e_{id} + (s-1)e_r T_2 + (s-1)(e_t+e_d a^n) T_1 \qquad (13)$$

At the end of 2*(2B/D)*(s-1) secs the leader in each segment has one data packet for that segment and now they have to relay this packet to the sink via the leaders in the neighboring segments. The process by which all the requisite packets reach the sink is shown in Fig. 3. These packets have to be then relayed to the sink. For this at first it is ascertained whether K is even or odd. If K is even then at first the leaders in the even segments are in the transmitting mode while those in the odd segments are in the receiving mode. The situation is exactly opposite when K is odd. Now so that packets can be correctly received it is required that the leaders transmit among themselves with $R_{radio}$ = 2b. This would ensure that connectivity is maintained. However collision of data packets can be avoided. Hence the value of 'd' used in (7) is going to be 2b. We illustrate a case when K=4. Each step like this is called a 'Transfer Round' and is depicted in Fig. 3 as a Step. Next leaders in the odd segments undertake the task of transmitting data while those in the even sections go into the receiving mode. This entire process comprising of various 'Transfer Rounds' by which all data packets are transferred to the sink is called a 'Transfer Cycle'. So that resulting clock drifts do not cause synchronization problems the same procedure as discussed earlier is adopted here as well. Now from the figure is can be easily deciphered that the entire duration of the Transfer Cycle is (2K-1)*(2B/D) secs. Once all the required data packets have been transferred by the leaders in a section to those in the adjacent segment these nodes go into the sleep state for the remainder of the data gathering cycle. For example: the leader in the final segment has only a single packet to transmit during a data gathering cycle and so after accomplishing this task it goes into the sleep state for the remaining part of the data gathering cycle to conserve energy.

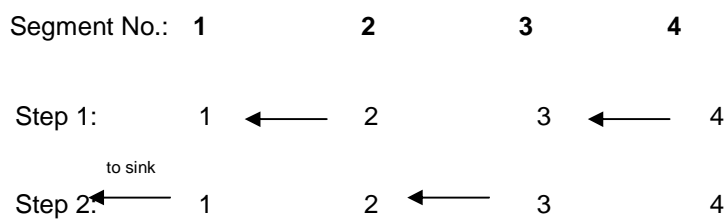

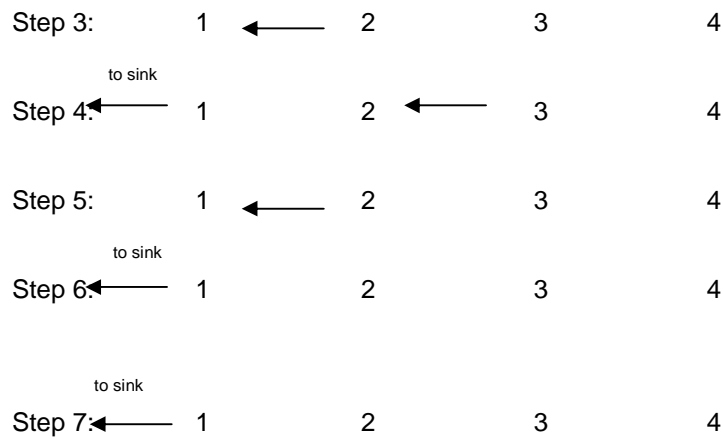

Fig.3. The Transfer Cycle

From the discussions above we can also calculate $T_{idlei1max}$.

$$T_{idlei1max} = 25(K+s-i-1) \quad (14)$$

Neglecting the time required for fusion of data packets the minimum duration of each data gathering cycle in ms in our case with the values considered above is thus given by,

$$T_{dmin} = T_{sense} + 100K + 50(s-1) \quad (15)$$

Further the time spent by an active node in the sleep state is given by,

$$T_{sleepa} = 100K + 50(s-2) + 25 \quad (16)$$

Similarly the minimum time spent by the leader in the sleep state is given by,

$$T_{sleeplmin} = 100K - 25(K-i+1) \quad (17)$$

The working of the entire network can be described in a compact manner with the help of flow diagrams. Following the steps highlighted in Figure 4 a node can easily determine its course of action. The flowchart can also be easily converted to any programmable logic. The flowchart demonstrates how a node can decide whether it will take part in data collection during a particular data gathering cycle. It also illustrates how a node can determine whether it is going to become a leader during a data gathering cycle or is going to simply act as an active node. The flowchart also indicates the steps to be followed by the leader as well as the active node.

**Figure 4: Flowchart**

A: Is My Serial Number Between $S_{no}$ and $S_{no} +s -1$?
B: Is (My Serial Number – $S_{no}$) = Number of data Gathering Cycles passed in a Round?
C: Is My Serial Number less than that of the LEADER?

D: Is the Value of K odd?

E: Is my Section Number odd?

F: Have I Transmitted K-i+1 Packets?

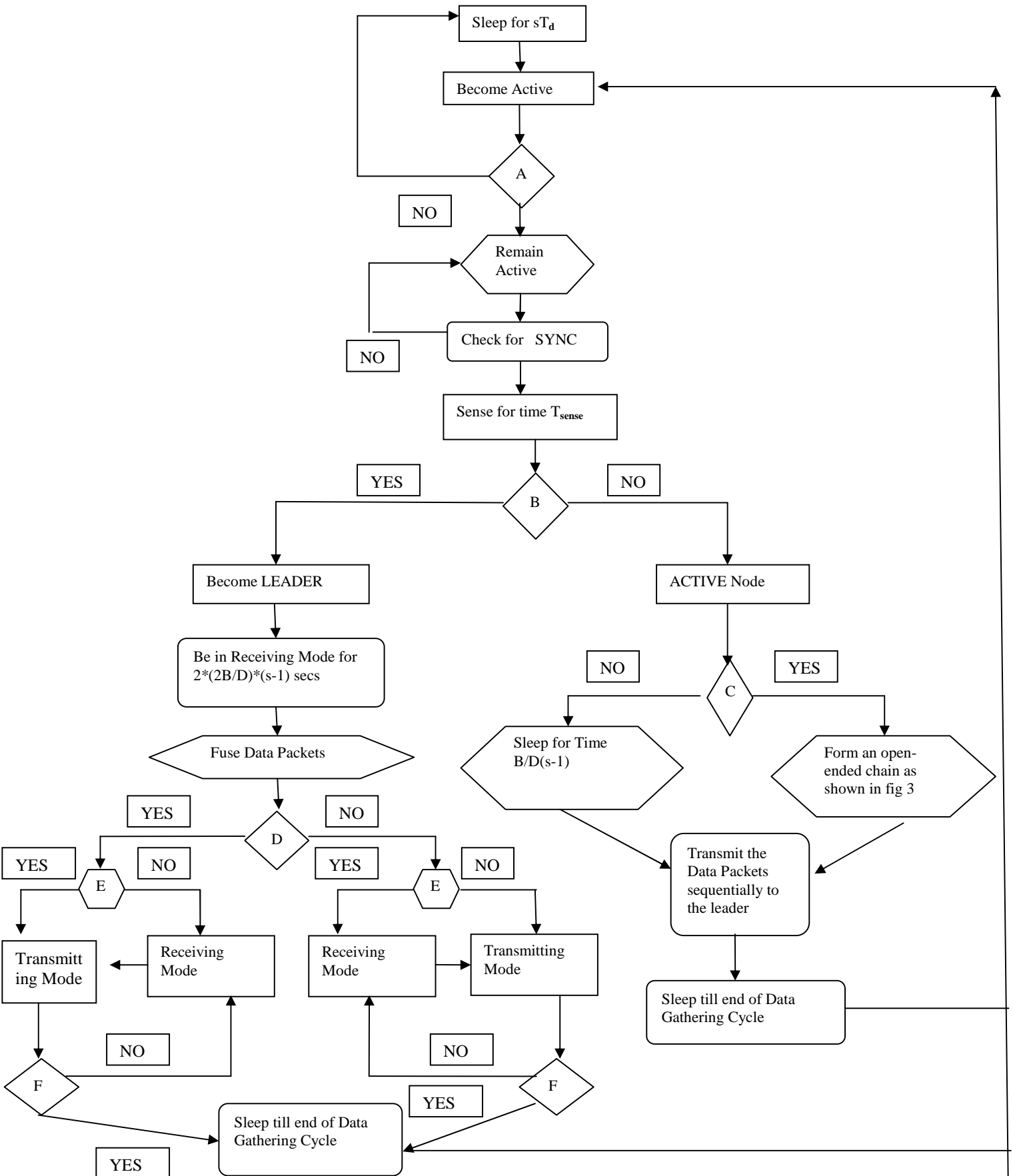

## VIII. FAULT TOLERANCE MANAGEMENT

In each data gathering cycle, in each section at any time 's' nodes are active. In case any of the 's' nodes fail, the chain breaks and they cannot relay the data to the leader of that section. Thus the network should be designed with a large number of redundancies to achieve fault tolerance and to maintain the desired network lifetime and coverage. The optimal number of sensor nodes required in each section is given by [12].Though the following equation has been developed for layered wireless sensor networks, the same concept can be applied here also, considering this network to have a single layered structure. The optimal number of redundant nodes required in each section is given by [12],

$$N = [Cd^2 (\Sigma n+1) \tau N_{active\_n} T_{totalsense}] / (E_{node} T)$$

(Where N is the number of redundant nodes to be deployed in each section, *c* is proportionality constant; its value mainly depends on the electromagnetic conditions in the atmosphere where the information is transmitted, *d* is the distance between two adjacent layers. *Σn* is the number of nodes from the outer layers which are connected to a node in Layer *n*, *τ* is the transmitting time for a node to transmit its sensed information, i.e., 1 packet, *Nactive_n* is the number of active nodes required in Layer *n*,*Ttotalsense* is the total monitoring period, i.e., the desired network lifetime, *Enode* is the total energy available for each node. All the sensor nodes are assumed to have the same initial total energy. *T* is the time period between two transmissions for the same node). Every node has a list of the N redundant nodes in its section.

The network we have proposed cannot have any packet drop due to collision owing to its design. If there is any packet drop it will be either due to node or link failure. In case a node 'a' fails to transmit a packet to its neighboring node 'b' in a transfer cycle for a maximum of 'm' attempts ('m' is taken as 10 in exponential backoff algorithm), then either the transmitting node 'a' or the receiving node 'b' or the link connecting 'a' and 'b' is at fault. In that case 'a' performs a self-check by transmitting a beacon packet to another neighboring node 'c'.
Case 1: If 'a' fails in this case too then it is at fault. Every node starts a timer from the time it is supposed to receive any data packet from its neighboring node in any transfer round. The timer interval is set to 2B/D secs i.e., twice the time interval needed to transfer a data packet successfully between 2 nodes. In case 'b' fails to receive any packet from 'a' within m*(2B/D) secs,then it concludes 'a' is at fault.
Case 2: If 'a' succeeds to send the beacon packet to 'c', then either 'b' or the link connecting 'a' and 'b' is at fault
Node 'b' in Case (1) and node 'a in Case (2) chooses the redundant node 'd' nearest to it and assigns it the serial number of the node that just failed. Then it also informs the other s-2 nodes active at that time of the change in serial number of the redundant node. Then the communication proceeds normally. For the remaining of that transfer round and till the failed node is repaired the node 'd' performs the functions of node 'a' in Case (1) and node 'b' in Case 2.

## IX. SIMULATION RESULTS

The simulations have been performed in NS2 on areas of different specifications. Extensive simulations are carried out in ns2 by changing the various parameters like network lifetime ($T_{life}$), initial energy of nodes ($E_o$) and the value of K on areas of different dimensions. However, only a few sample results are listed here. We initially consider area of same dimension as in Section IV. Here a=60, b=10 and K=10. A comparative study between the value of s and the coverage fraction (β) is performed and this is plotted in Figure 5.

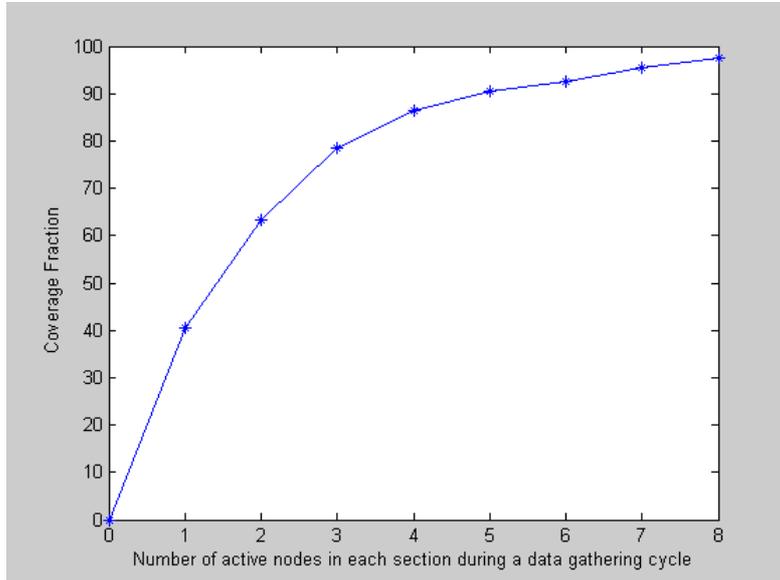

**Figure 5: Coverage Fraction**

We state the values of the various parameters used for simulation. In this simulation we take the data rate as 20 kbps, B=512 bits and also assume $T_1=T_2$. The values of the different parameters used in the simulation are similar to the one taken in [2] obtained by modifying the parameters in a meaningful way for our scheme. We take $e_t$=1024μJ/s, $e_r$=819.2μJ/s, $e_d$=2048 nJ /m$^2$/s, and $e_{id}$ = 409.6μJ/s. There is also cost for sensing the surroundings. We take for simulation purposes this energy as $e_{sense}$ = 81.2μJ/s and $T_{sense}$ was taken to be 3 seconds.

The value of $T_d$ was chosen to be 4.4 seconds because $T_{dmin}$ =4.35 seconds. In TABLE I an estimate of the number of nodes to be deployed in the area with a=60m, b=10m, K=10 and $E_o$ = 1000J and other parameters as mentioned above. We assume that β =90% and $T_{life}$ = 5 years is given. The $R_{radio}$ for the first part of the data gathering cycle was taken as 10m and 20m for the second part of the data gathering cycle. The path loss is taken as 2.The lifetime that is actually achieved when simulation is done with ns2 is 4.58 years.

In Fig 6, the amount of energy dissipated by the nodes in the various segments during each data gathering cycle for different packet sizes and data rate is shown which gives an indication as to why changing number of nodes needs to be deployed in the various sections. It also gives a comparison between the analytical results and those obtained in the real world scenario (simulated in ns2) for the energy consumption of the active nodes considering different packet sizes and data rates. The difference in the energy consumption is natural when this analytical model is being incorporated in a real network where we have to take in consideration the ACK packets, packet drop and retransmission due to accidental packets drops etc. and such network features which are not possible to incorporate in the analytical model.

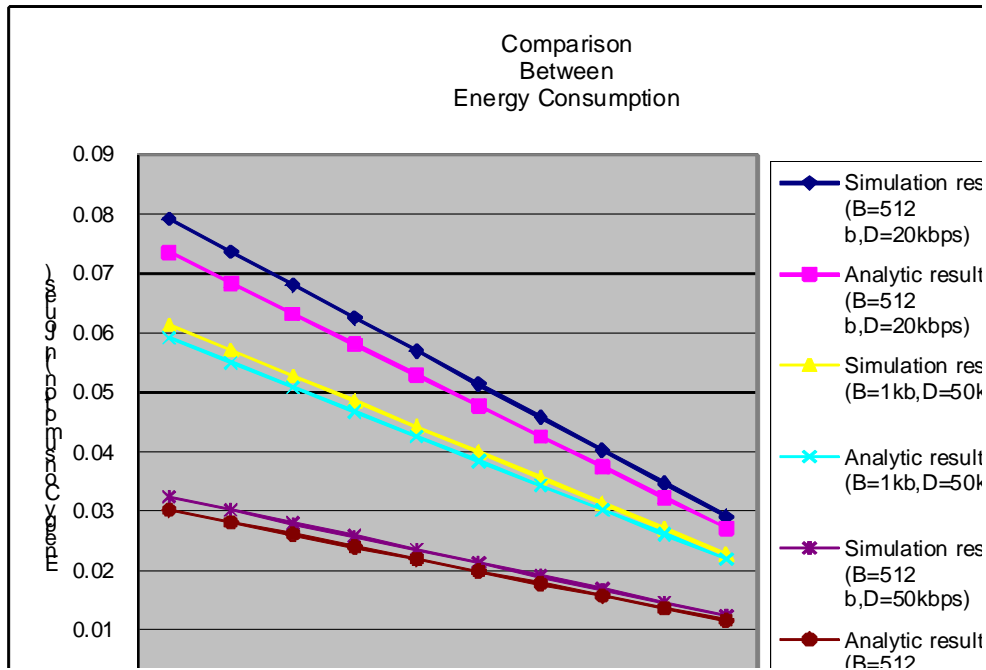

**Figure 6: Comparison of Energy consumption of active nodes**

In Table I an estimate of the number of nodes to be deployed in the area with a=60m, b=10m, K=10, $E_o$ = 1000J, β =90% and $T_{life}$ = 5 years is given

**TABLE I: Segment vs. number of nodes**

| Segment number | Number of Nodes |
|---|---|
| 1 | 2640 |
| 2 | 2455 |
| 3 | 2270 |
| 4 | 2085 |
| 5 | 1900 |
| 6 | 1715 |
| 7 | 1530 |
| 8 | 1345 |
| 9 | 1160 |
| 10 | 975 |

A comparative study between the initial energy of the nodes and the total number of nodes required to be deployed for different data rate and packet sizes for a given network lifetime of 5 yrs was also done and the result is shown in Table II.

**TABLE II: Initial Energy vs. Total number of nodes**

| B=512 bytes,D=20kbps | |
|---|---|
| Initial Energy (J) | Total number of Nodes |
| 1000 | 18075 |
| 2000 | 9050 |
| 3000 | 6040 |
| 4000 | 4435 |
| B=1024 bytes, D=50kbps | |
| Initial Energy (J) | Total number of Nodes |
| 1000 | 14540 |
| 2000 | 7280 |
| 3000 | 4865 |
| 4000 | 3655 |

Assuming $T_{life}$ to be 5 years the total number of nodes to be deployed in area under surveillance was calculated. As expected with an increase in the initial energy possessed by the nodes number of nodes are required to be deployed for achieving a network lifetime is found to decrease. When the initial energy possessed by the nodes is 1000J the total number of nodes required is 18075 and when the initial energy is 4000J the total number of nodes to be deployed is found to be 4435 given the data rate is 20kbps and packet size is 512 bytes. For the area considered, an analysis of the projected lifetime and the obtained lifetime is given in Table III. Here we consider $E_o$ = 1000J for simulation.

**TABLE III: Projected Life Time vs. Obtained Life Time**

| B=512 bytes, D=20kbps, Initiatial Energy=1000 J | |
|---|---|
| PROJECTED LIFE TIME (YEARS) | OBTAINED LIFE TIME (YEARS) |
| 5 | 4.65 |
| 10 | 9.29 |
| 25 | 23.22 |
| 50 | 46.43 |

## X. CONCLUSION

The scheme considered in this paper not only ensures that nodes dissipate minimum amount of energy during a data gathering cycle, but also takes into account that complete energy utilization occurs thereby increasing network lifetime. Moreover the radio range is chosen in a manner that the full connectivity among nodes in adjacent segments is always maintained independent of their distribution in a segment. Apart from all this an estimate of the number of nodes to be deployed in order to achieve a required network lifetime is also provided. The simulation results also help to understand and appreciate the facts stated in the paper. The paper also discusses fault tolerance management in details to ensure proper functioning of the network in spite of node or link failure.